\documentclass[12pt]{article}
\usepackage{times}
\usepackage{geometry}
\usepackage[utf8]{inputenc}
\usepackage{enumitem,amssymb}
\usepackage{ragged2e}
\newlist{thematic}{itemize}{8}
\setlist[thematic]{label=$\square$}
\usepackage{pifont}
\newcommand{\cmark}{\ding{51}}%
\newcommand{\done}{\rlap{$\square$}{\raisebox{2pt}{\large\hspace{1pt}\cmark}}%
\hspace{-2.5pt}}

\newcommand{\hi}{H{\sc i}~}
\newcommand{\HI}{H{\sc i}}

\usepackage{graphicx}
\usepackage{natbib}
\usepackage{relsize}

\usepackage{titlesec}
\titlelabel{\thetitle.  }

\titleformat*{\section}{\large\bfseries}
\titleformat*{\subsection}{\normalsize\bfseries}

\newcommand\araa{ARA\&A}
\newcommand\apj{ApJ}
\newcommand\apjl{ApJ}
\newcommand\apjs{ApJS}
\newcommand\aap{A\&A}
\newcommand\mnras{MNRAS}
\newcommand\prl{Phys.~Rev.~Lett.}
\newcommand\procspie{Proc.~SPIE}

\newcommand\ion[2]{\mbox{#1$\;${\smaller\rmfamily\uppercase\expandafter{\romannumeral #2 \relax}}}}

\begin{document}
\raggedright
\large
Astro2020 Science White Paper \linebreak

\huge
Magnetic Fields and Polarization in the Diffuse
Interstellar Medium\linebreak
\normalsize

\noindent \textbf{Thematic Areas:} \hspace*{60pt} $\square$ Planetary Systems \hspace*{10pt} $\done$ Star and Planet Formation \hspace*{20pt}\linebreak
$\square$ Formation and Evolution of Compact Objects \hspace*{31pt} $\square$ Cosmology and Fundamental Physics \linebreak
  $\square$  Stars and Stellar Evolution \hspace*{1pt} $\square$ Resolved Stellar Populations and their Environments \hspace*{40pt} \linebreak
  $\square$    Galaxy Evolution   \hspace*{45pt} $\square$             Multi-Messenger Astronomy and Astrophysics \hspace*{65pt} \linebreak
  
\textbf{Principal Author:}

Name: S. E. Clark
 \linebreak						
Institution: Institute for Advanced Study
 \linebreak
Email: seclark@ias.edu
 \linebreak
 
\textbf{Co-authors:} 

Carl Heiles, University of California at Berkeley, heiles@astro.berkeley.edu\\
Tim Robishaw, Dominion Radio Astrophysical Observatory, tim.robishaw+drao@gmail.com
  \linebreak

\textbf{Abstract:} 

Magnetism is one of the most important forces on the interstellar medium (ISM), anisotropically regulating the structure and star formation that drive galactic evolution. Recent high dynamic range observations of diffuse gas and molecular clouds have revealed new links between interstellar structures and the ambient magnetic field. ISM morphology encodes rich physical information, but deciphering it requires high-resolution measurements of the magnetic field: linear polarization of starlight and dust emission, and Zeeman splitting. These measure different components of the magnetic field, and crucially, Zeeman splitting is the only way to directly measure the field strength in the ISM. We advocate a statistically meaningful survey of magnetic field strengths using the 21-cm line in absorption, as well as an observational test of the link between structure formation and field strength using the 21-cm line in emission. Finally, we report on the serendipitous discovery of linear polarization of the 21-cm line, which demands both theoretical and observational follow-up. \linebreak

\textbf{Related Astro2020 white papers:}\\
\textit{Twelve decades: Probing the interstellar medium from kiloparsec to sub-AU scales}.\\ 
Principal Author: Stinebring\\
\vspace{0.1in}
\textit{Studying magnetic fields in star formation and the turbulent interstellar medium}.\\ Principal Author: Fissel\\

\thispagestyle{empty}
\clearpage
\pagenumbering{arabic} 

\justify
\section{Introduction}

The `neutral' interstellar medium (ISM) is ionized to a small degree,
$\sim10^{-4}$, enough for the gas to be coupled to the magnetic field. Magnetic forces are fully competitive with gravity, turbulent
pressure, and cosmic ray pressure in determining gas dynamics. 
Magnetic forces are anisotropic and are transmitted from the rare electrons to
the plentiful neutral Hydrogen (\HI) by collisions and cosmic-ray coupling,
both of which become ineffective at small length scales and high volume
densities \citep{hennebellef11, mckee07, snez12}. These magnetically-related phenomena not only make the
ISM a fascinating entity in itself, but also lead to the multifaceted
process of star formation---and, by extension, galactic and cosmological
evolution.

\section{Observing Magnetic Fields in the ISM}
Invisible interstellar magnetic fields are revealed by the linear polarization of starlight and of infrared dust emission, which are produced by
magnetically-aligned dust grains; these polarizations trace the
orientation of $B_\perp$, the plane-of-the-sky component, but provide no
information on magnetic field strength. Zeeman splitting produces
circular polarization that provides the magnetic field strength and
direction; the splitting is very small compared to the line width, which
makes the measurements sensitivity limited.

\subsection{Tracing the Magnetic Field with \hi Fibers} \label{susan}

\begin{figure}[h!]
\begin{center}
\vspace*{-.35in}
\includegraphics[scale=0.45]{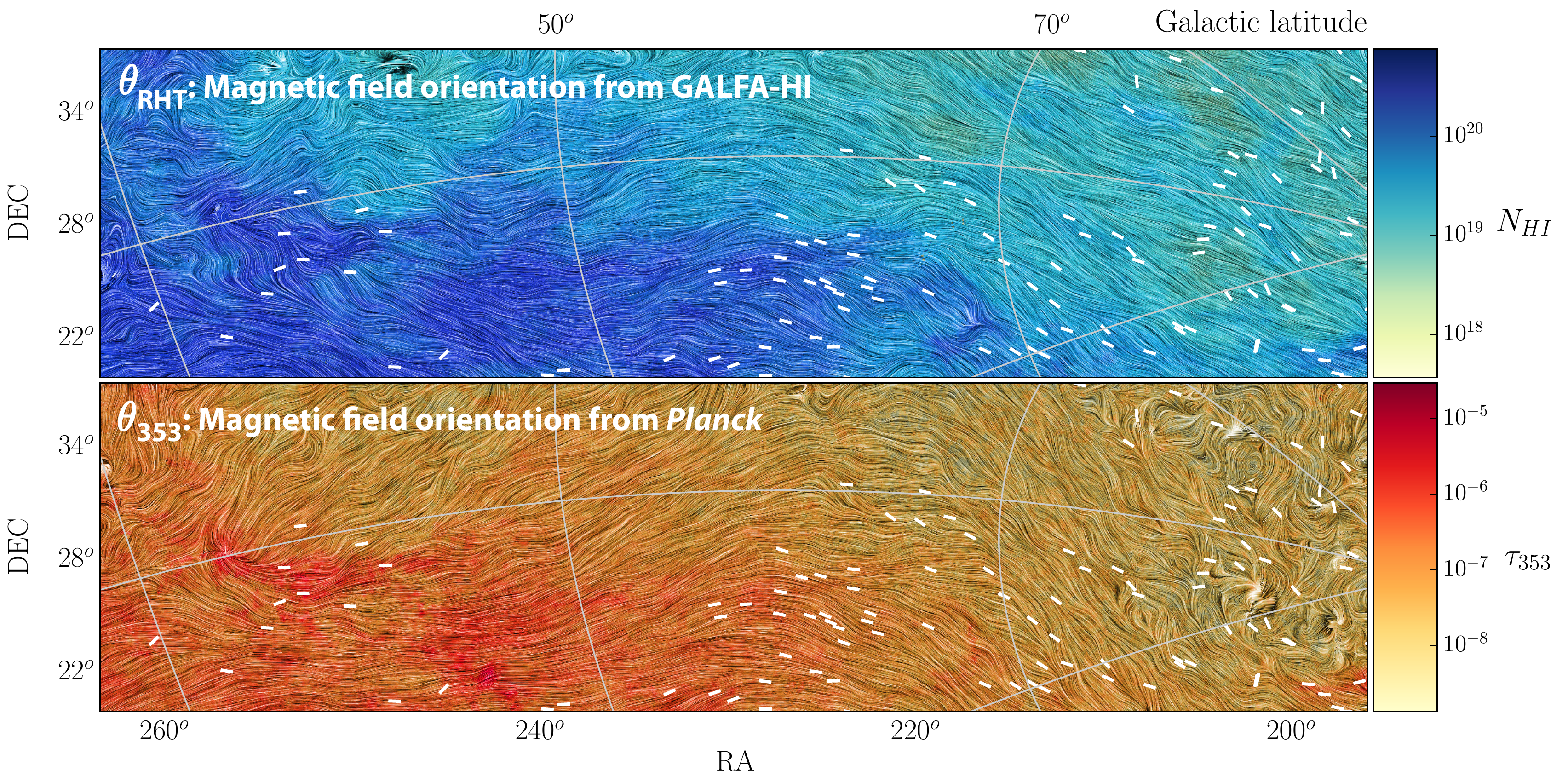}                                         
\end{center}
\vspace{-4ex}
\caption{\footnotesize Top: texture shows the orientation of \hi structures. Bottom: texture shows the plane-of-sky magnetic field orientation implied from \textit{Planck} 353 GHz polarized emission. White: plane-of-sky magnetic field orientation from optical starlight polarization measurements. Linear \hi structures are well aligned with the ambient magnetic field. \textit{Figure reproduced from \citet{Clark:2015}.
\label{fibers}}}
\vspace*{-.1in}
\end{figure}

 The neutral ISM is textured by the ambient magnetic field (see Figure
 \ref{fibers}). Sensitive, high dynamic range observations of the 21-cm
 line in emission reveal that high aspect ratio (up to 100:1) \hi
 structures (``\hi fibers") thread the diffuse ISM and trace the local
 magnetic field orientation as probed by optical starlight polarization
 \citep{Clark:2014} and \textit{Planck} polarized dust emission
 \citep{Clark:2015, Kalberla:2016}. Magnetically aligned fibers are a fairly common feature of \hi emission. The \hi fibers are preferentially cold neutral medium \citep[CNM;][]{Clark:2014,
   Clark:2019, Kalberla:2016} and thus bear
 important similarities to slender, magnetically dominated \hi
 self-absorption structures in the Galactic Plane
 \citep{McClureGriffiths:2006}.
 
Information about the magnetic field is thus encoded in the
\textit{morphology} of \hi emission and absorption. The shape
information in the ISM carries important physical meaning, but decoding
it requires 1) high dynamic range observations and 2) algorithms for
quantifying the spatial information. The recent Galactic Arecibo L-Band Feed Array Survey
(GALFA-\HI) mapped the 21-cm line with unprecedented dynamic range over the Arecibo sky
\citep[4$'$ spatial resolution over 4 sr;][]{Peek:2018}. The \HI4PI Survey \citep{HI4PI:2016} achieved an
all-sky map at $16'$ by combining the Effelsberg-Bonn \hi Survey of the northern sky \citep[EBHIS;][]{Winkel:2016} with the Parkes Galactic All-Sky Survey in the south \citep[GASS;][]{McClure-Griffiths:2009}. Other diffuse ISM tracers, particularly those
that yield direct information on the magnetic field, have not been
mapped with comparable resolution. The \textit{Planck} satellite was a
major leap forward for far infrared polarimetry, but \textit{Planck} achieved
only $\sim 60'$ resolution observations of polarized dust emission in
the diffuse ISM \citep{Planck:2018XII}. An all-sky survey of polarized dust emission like the
one proposed for the probe class concept PICO \citep[Probe of Inflation and Cosmic Origins;][]{Hanany:2019} would revolutionize the study of
magnetic structure in the ISM. The second point---algorithms---is a need
for methods that quantify the spatial distribution of emission
data. The rich physical information in the morphology of the diffuse ISM
mirrors the spectacular advances of the last decade in the study of
molecular cloud structure, in which \textit{Herschel} observations
revealed that molecular clouds are universally filamentary
\citep{Molinari:2010, Arzoumanian:2011, Palmeirim:2013}. These
observations motivate innovative techniques for quantifying spatial
structure in the ISM, and its relationship with the local magnetic field
\citep[e.g.][]{Soler:2013}.

\begin{figure}[h!]
\begin{center}
\vspace*{-.3in}
\includegraphics[scale=0.7, bb=327 502 521 695, clip]{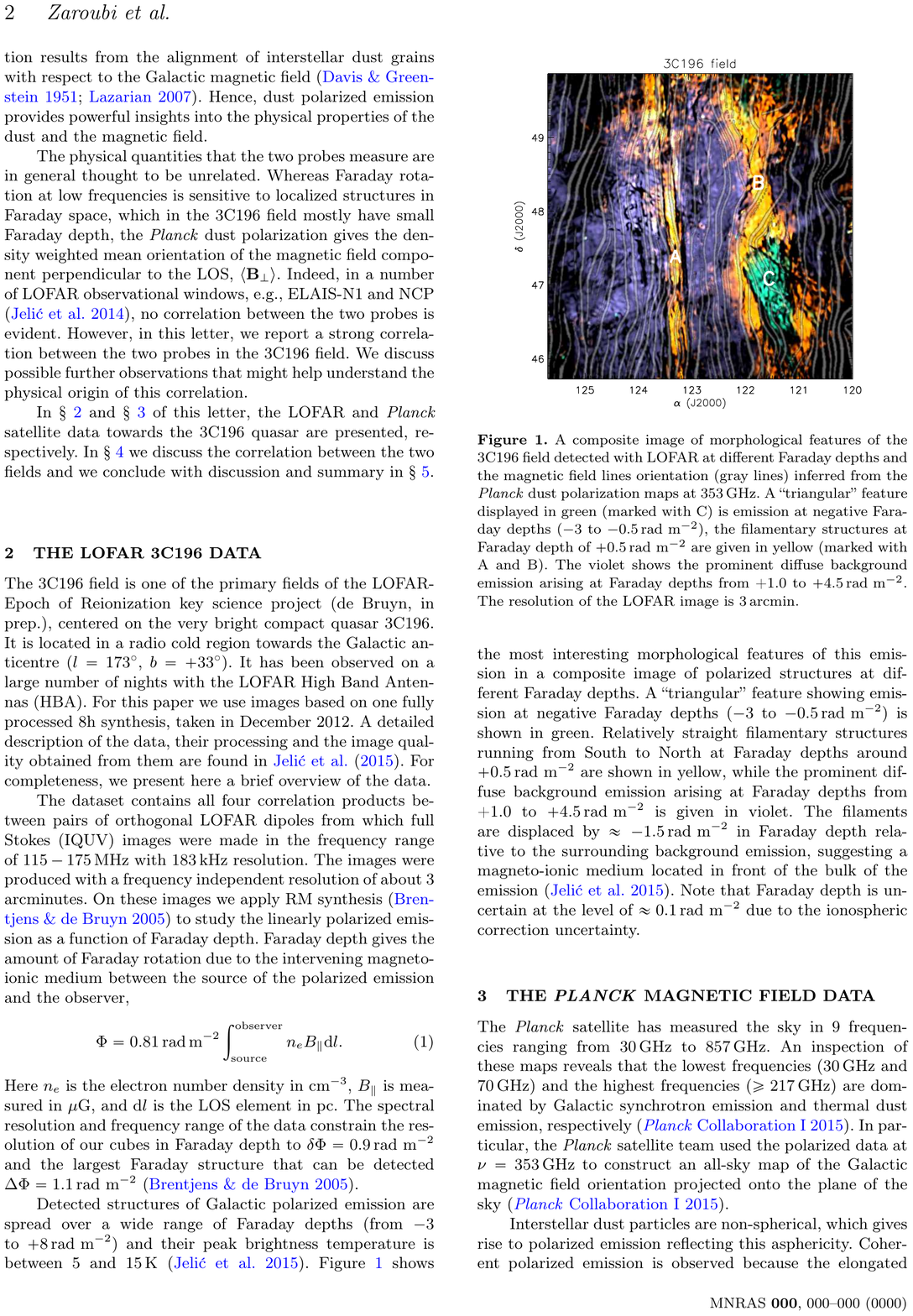}                       
\includegraphics[scale=0.7, bb=74 561 301 741,clip]{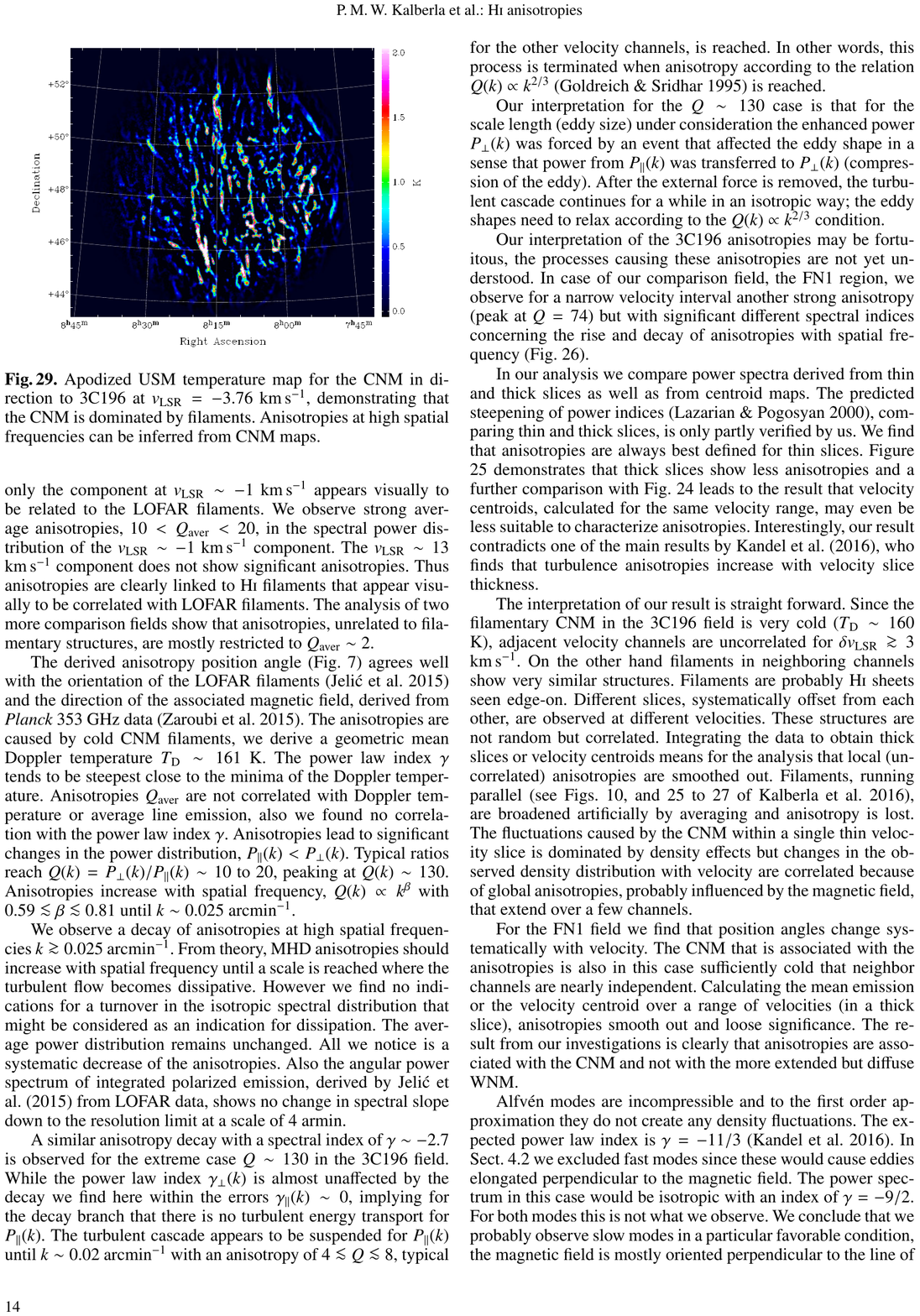}
\end{center}
\vspace{-4ex}
\caption{\footnotesize {Left:} 3C196 field showing distinct features at three different Faraday depths (purple, yellow, green), and \textit{Planck} 353 GHz magnetic field orientation (gray contours). \textit{Figure reproduced from \citet{Zaroubi:2015}.} {Right:} Anisotropic \hi structures in the same field. \textit{Figure reproduced from \citet{Kalberla/Kerp2016}.} \label{fibers196}}
\vspace*{-.1in}
\end{figure}

Disparate observations show an intriguingly similar phenomenon: highly
anisotropic, magnetically aligned structures on small scales in the
ISM. A striking example in the magneto-ionic medium is the observation
of high aspect ratio linear polarization structures with The Low-Frequency Array
\citep[LOFAR; ][]{Jelic:2015}, see Figure \ref{fibers196}, left panel. These are
straight, and well aligned both with the local magnetic field as probed
by polarized dust emission \citep{Zaroubi:2015, Jelic:2018} and with
filamentary \hi structures \citep{Kalberla/Kerp2016}. The implication is that the different phases of the ISM know about one
another and/or a shared magnetic field. Does this knowledge extend down to the AU-size scales probed by pulsar scintillation \citep{Stinebring:2019}? Are these structures related to a
particular geometry, such as the compressed shell of the Local Bubble
\citep[e.g.][]{Lallement:2014}, or are they a signature of similar diffuse structure formation processes across phases? Parsing the relationship
between the magnetic field and the phases of the ISM requires
high-resolution observations of magnetic field structure in the
molecular gas and neutral medium (Zeeman splitting,
polarized dust emission, starlight polarization), as well as the warm ionized
medium (Faraday rotated diffuse synchrotron polarization).

The discovery that \hi structure carries information about the
local magnetic field orientation has tantalizing implications for
mapping the magnetic field structure in three dimensions. \ion{H}{1}
emission is measured in position-position-velocity space: the third
dimension available to spectral line observations is not spatial, but
the Doppler-shifted emission frequency along the line of sight. With
stellar photometry of millions of stars, huge progress has been made
recently in mapping dust reddening in three spatial dimensions
\citep{Green:2018}. With the advent of \textit{Gaia}, the accuracy of
stellar distance measurements has vastly improved.  Still lacking are
ample starlight polarization measurements, both in the optical, for
studying the diffuse ISM, and in the infrared, for extincted sightlines
in the Galactic plane \citep[e.g. GPIPS;][]{GPIPS:2012}. The Polar-Areas Stellar Imaging in Polarization High-Accuracy Experiment will provide a major advance, measuring linear polarization
toward millions of stars at high Galactic latitudes
\citep[PASIPHAE;][]{PASIPHAE:2018}. 
A statistical sample of Zeeman splitting observations (\S \ref{absorption}) would
provide the crucial missing component: the magnetic field {\it 
strength}.

In addition to being a fascinating physical system in its own right, the
diffuse, magnetized ISM is also a formidable foreground for cosmological
observations. In particular, the much-sought-after inflationary
gravitational wave $B$-mode polarization signal in the cosmic microwave
background (CMB) is buried under much brighter polarized Galactic
emission at all wavelengths \citep[][]{BicepPlanck:2015,
  Kogut:2016}. All of the
observations discussed here -- \hi emission structure, diffuse polarized
dust emission, starlight polarization data, Zeeman measurements of the
magnetic field strength -- will aid CMB foreground removal
\citep{Tassis:2015, Clark:2018, Hensley:2018}.

\subsection{Building
an Unbiased Sample of Magnetic Field Strengths 
in the CNM Using the \hi Line in Absorption} \label{absorption}

The line-of-sight component of the magnetic field strength ($B_{||}$) in
the CNM can be probed via the absorption of
continuum emission from compact background sources, which are
distributed randomly on the sky. This method was successfully used by
\citet{heilest04,heilest05} at the Arecibo telescope to make the
unprecedented ``Millennium'' survey of the CNM towards 42 radio-loud
sources. The unique products of such observations include not only the
field strength, but also accurate kinetic temperatures---and, from the
line width, the degree of turbulence and whether it is supersonic and/or
super-Alfv\'{e}nic. This combination of physical properties is essential for 
understanding the magneto-gas dynamics and can be obtained by no other 
observational technique.

\begin{figure}[h!]
\begin{center}
\vspace*{-.1in}
\includegraphics[scale=0.4, trim={0 0 0cm 0cm},clip]{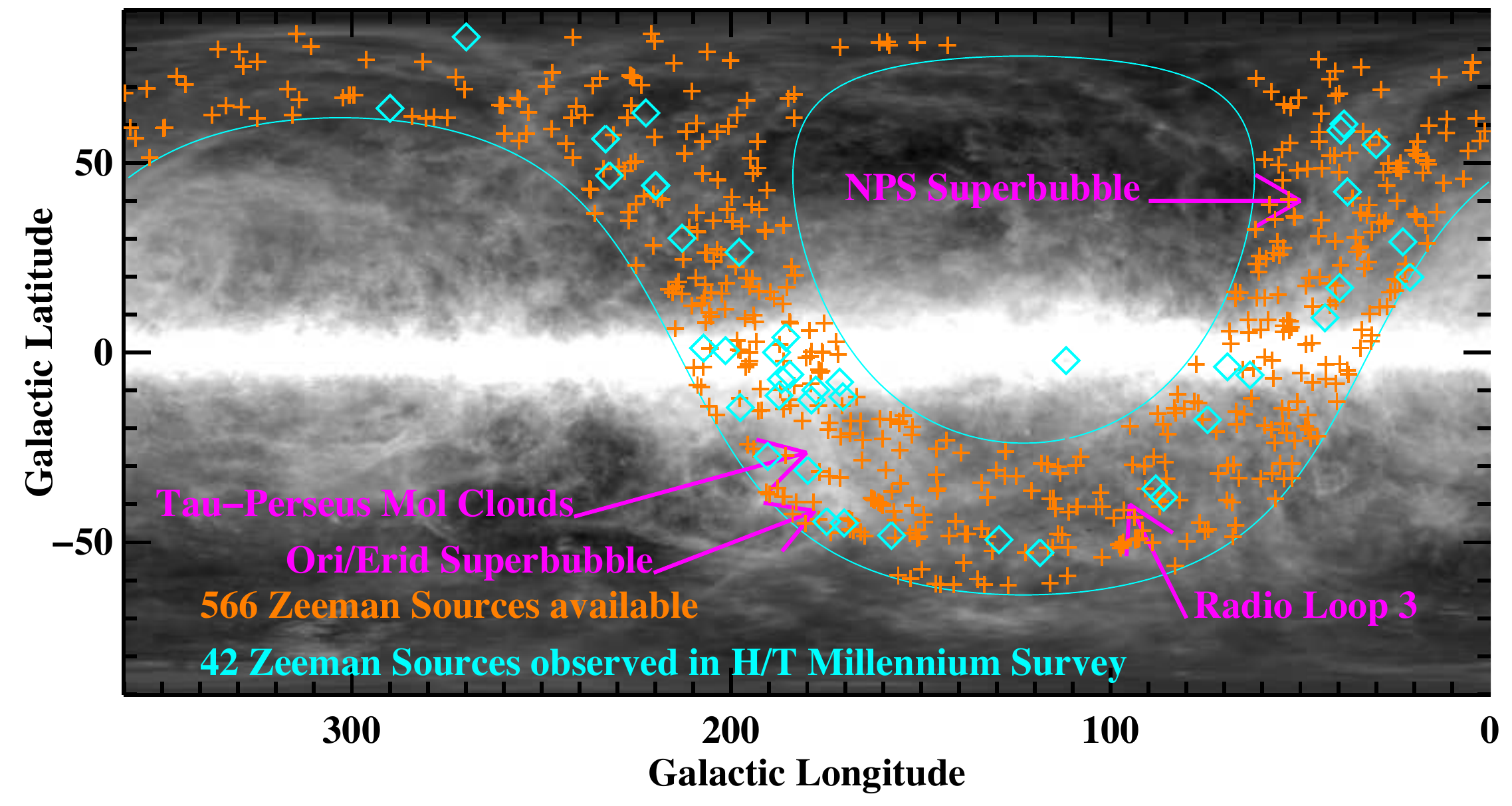}
\includegraphics[scale=0.4, trim={0 0 0cm 0cm},clip]{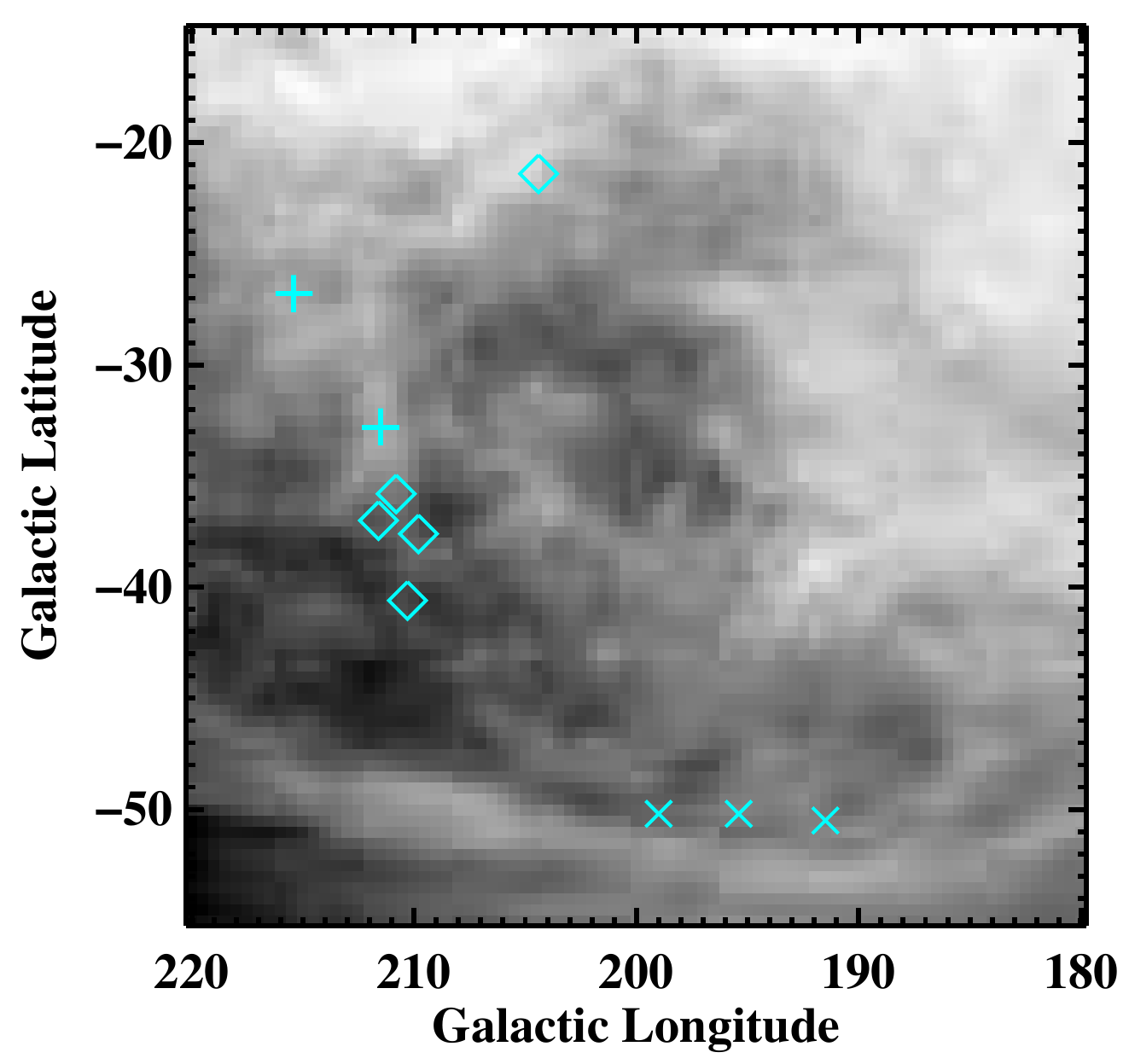}
\end{center}
\vspace{-4ex}
\caption{\footnotesize {\it Left:} the 42 Millennium survey sources
  (cyan diamonds) and the 566 sources exceeding 1 Jy (orange crosses)
  within the Arecibo declination range (cyan lines), superposed on a
  grey-scale image of \HI. {\it Right:} Magnetic field signs from
  Zeeman splitting of \hi emission lines. Pluses are positive,
  crosses negative, and diamonds are nondetections, superposed on a
  grey-scale image of the \hi line in a portion of the
  Orion/Eridanus superbubble. $|B_{||}|$ ranges from 4 to 11
  $\mu$Gauss. \label{zmnsrcs2panel}}
 \vspace*{-.1in}
\end{figure}

While these results are important and widely referenced, the number of
actual detections is, from a statistical standpoint, pitifully small. The cyan diamonds in the left panel of Figure \ref{zmnsrcs2panel} show the 42 sources. These constitute about 7.5\% of the total number of sources within the Arecibo declination range that exceed the
minimum useful flux density of 1 Jy. The Arecibo sky contains four important
and distinguishable interstellar structures: Perseus, the Orion/Eridanus
superbubble, the North Polar Spur superbubble (a.k.a.\ the Radio Loop 1
superbubble), and the Radio Loop 3 superbubble. Each of these entities
is sampled by only a handful of sources. This is simply inadequate to
develop a statistically reliable picture of the magnetic field and its
fluctuation within each structure.

\subsection{Sampling Magnetic Field Strengths in Shocks and Filaments Using the \hi Line 
in Emission}

\citet{heiles89} measured Zeeman splitting of Galactic \hi in
emission, targeting compressed supershell walls and filaments. Figure
\ref{zmnsrcs2panel} (right panel) shows the results for
a portion of the Orion/Eridanus superbubble. Three things stand out from
this image: 1. The \hi is swept up from the interior into shell walls, which
  themselves have considerable spatial structure that sometimes looks
  filamentary. 2. The observational sampling is totally inadequate relative to the
  structural details of the \HI. 3. There is an apparent large-scale field reversal across the superbubble. However, the sampling is so coarse that no statistically firm statement about field reversal can be made---at {\it any} scale.
 The advantage of emission over absorption measurements is that one can
choose interesting positions based on ISM structure; one is
not limited to the arbitrary positions of background continuum sources. \citet{heiles89} sampled 73 positions in superbubble walls, which are compressed by shocks, and found that the measured field strengths are about twice those of the randomly-selected positions of \S \ref{absorption}.
These emission measurements were all made with the Hat Creek 85-foot telescope, which collapsed during a major windstorm in January 1993. Robishaw and Heiles are initiating collaborative development of instrumentation and techniques for the DRAO Galt 26-m, the Parkes 64-m, and the Effelsberg 100-m telescopes to expand the coverage of emission measurements. 

\section{A Serendipitous Discovery: Linear Polarization of the 21-cm Line}
\label{linpol}

For technical reasons we wanted to observe a set of test sources known
to be unpolarized. We expected the 21-cm line to be unpolarized because
its 2 level system is populated overwhelmingly by collisions.  We used
the Heiles \& Troland
Millennium survey (\citeyear{2004ApJS..151..271H}) data because of their long
integration times and careful polarization calibration.  We studied a
subset of 18 sources and found, much to our surprise, that 13 of these
18 sources exhibit detectable linear polarization at levels 0.14\% to
0.35\%. The linear polarization appears to be astrophysical, and we are currently exerting considerable effort to confirm or refute the finding.

\begin{figure}[t!]
\begin{center}
\vspace*{-.4in}
\includegraphics[width=3.2in, height=2.6in] {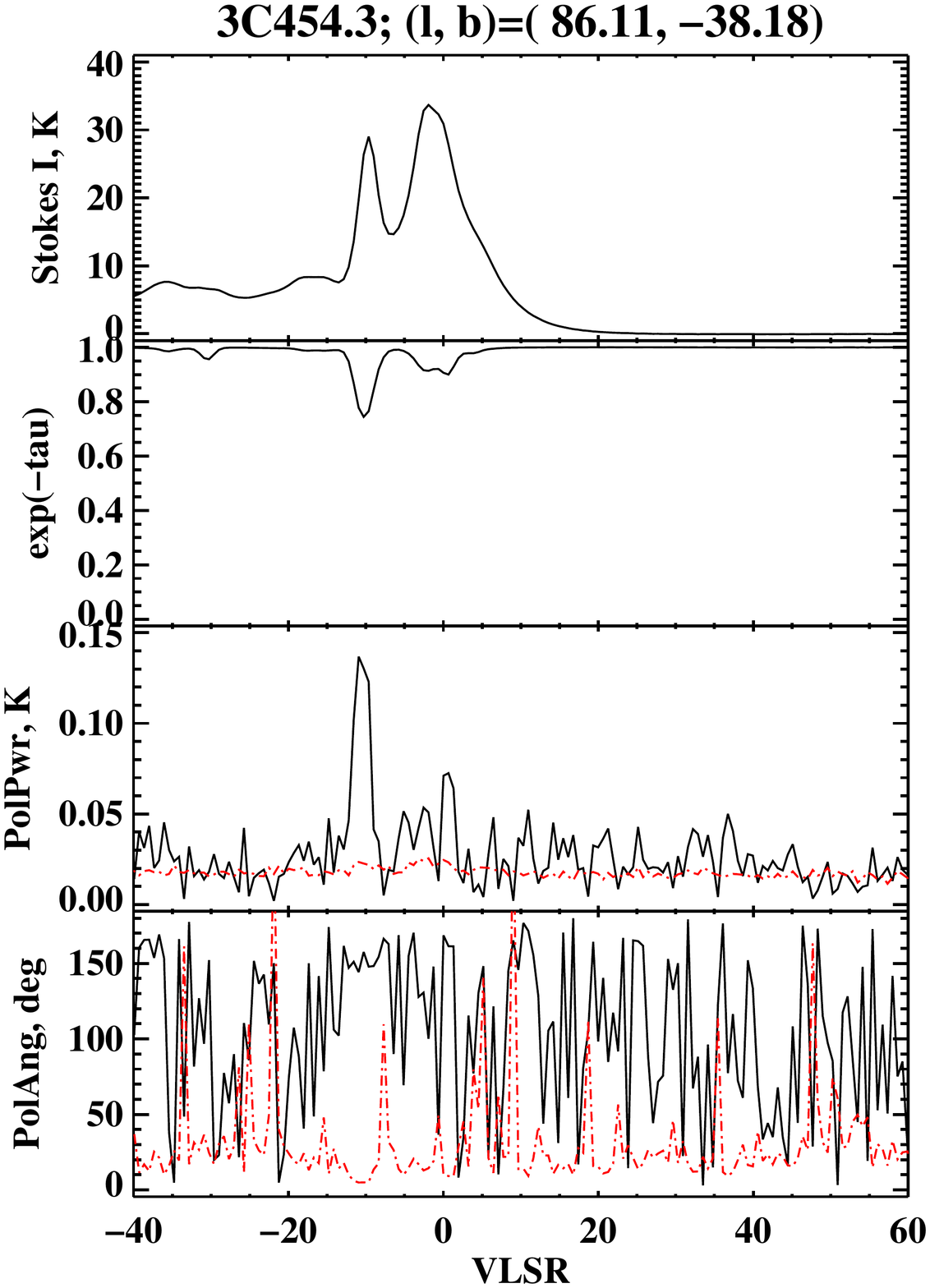}
\includegraphics[width=3.2in, height=2.6in] {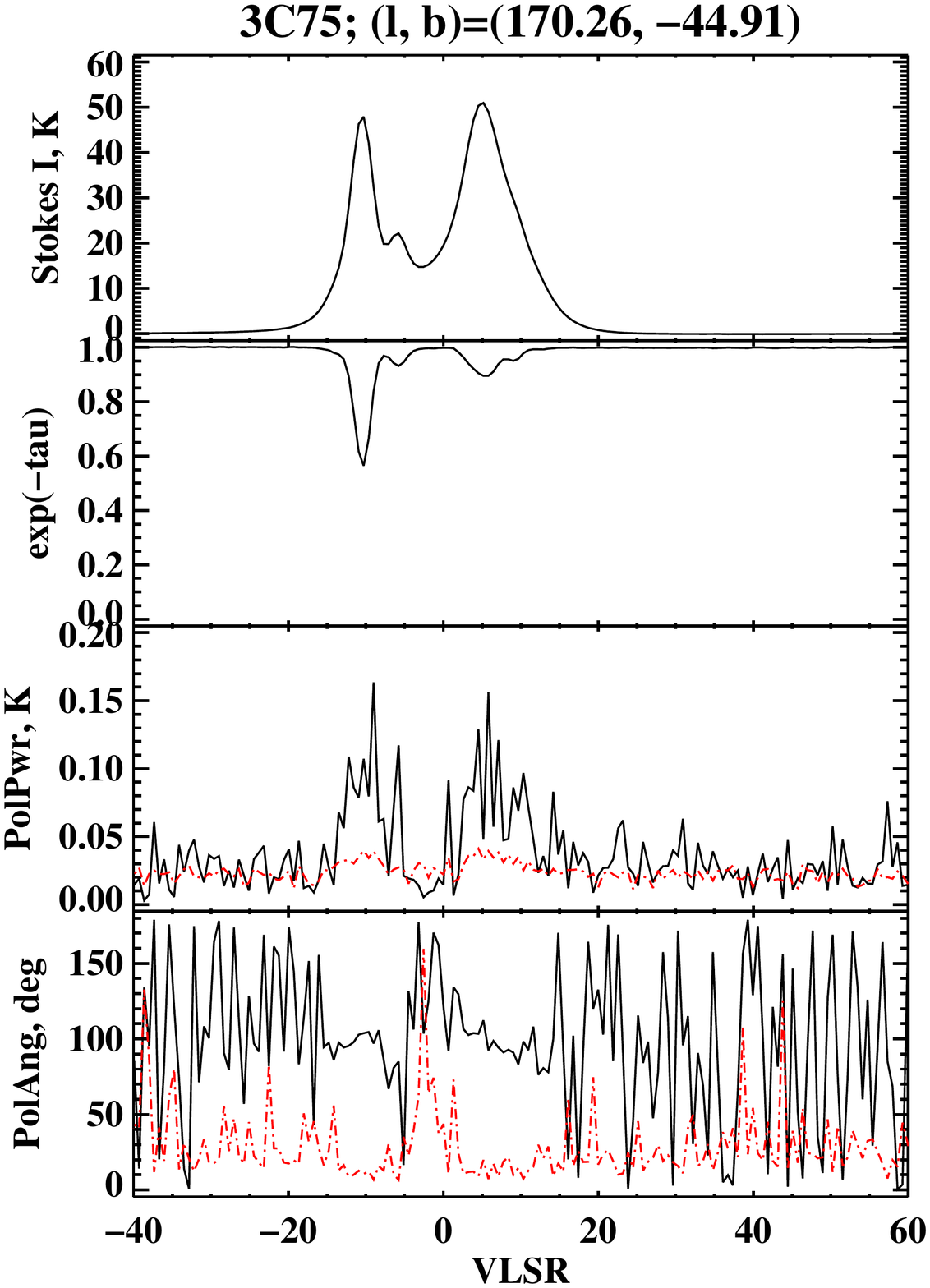}
\end{center}
\vspace*{-2em}\caption{\footnotesize Each of the four panels shows: top,
  Stokes I emission spectrum; next, the optical depth spectrum
  $\exp(-\tau)$; next, the total linear polarization power $(Q^2 +
  U^2)^{1/2}$; bottom, the position angle of linear polarization $\frac{1}{2}{\rm
    atan}(Q/U)$. The solid (black) line is the spectral quantity and the
  red (dashed) line its statistical uncertainty. \label{fig:pol_panels}}
  \vspace*{-.1in}
\end{figure}

Figure \ref{fig:pol_panels} shows two examples, 3C75 and 3C454.3. The
top panel exhibits the emission line and the absorption line.  The
bottom panels exhibit the linearly polarized power and the position
angle; the black lines are the profiles and the red their uncertainties.
Figure \ref{fig:pol_panels} shows that the polarization is not
exclusively associated with either the absorption profiles (the CNM) or
the emission profiles (both the warm neutral medium and the CNM). How can the collisionally-dominated two-level system show linear
polarization? While we idealize the hyperfine levels of the \ion{H}{1}
atom as a two-level system, this is not strictly true. Ly-$\alpha$
radiation is an important populating agent for the 21-cm levels
\citep{Murray:2014}, and could conceivably differentially populate the
magnetic sublevels, particularly if it is anisotropic in either physical
or frequency space. Indeed, we were surprised to learn that linear
polarization of the 21-cm line was predicted by
\citet{Yan/Lazarian2007}.  We speculate that the spatial anisotropies of
\S \ref {susan} are involved with the 21-cm line polarization.

\section {Instrumental Requirements for \S \ref{absorption} and 
\S \ref{linpol}}

Measuring circular polarization (\S \ref{absorption}) and linear
polarization (\S \ref{linpol}) requires obtaining not only the four
on-source \hi Stokes parameter spectra, but also the emission
spectra that would be observed if the source were absent. This requires
sensitive and accurate measurements of the \hi profile, which can
only be done with a large single dish telescope; interferometric arrays cannot
measure the spatially extended emission spectra with high sensitivity. 
Arecibo could supply more than 500 additional sources for the expenditure of $\sim 3000$ hours of telescope time using its new phased-array ALPACA feed, which allows 39 off-source emission spectra and the on-source spectrum to be measured simultaneously \citep{ALPACA:2016}.\\

\noindent The next decade is poised to dramatically advance our understanding of interstellar magnetism, one of the primary forces acting on the Galaxy. This will be achieved with an investment in polarimetry: surveys of polarized dust emission, starlight polarization, and single-dish measurements of Zeeman splitting.

\pagebreak

\newcommand{\noopsort}[1]{}

\end{document}